\documentclass[aps,prd,preprint,showpacs,superscriptaddress]{revtex4-1}
\usepackage{amsmath}
\usepackage{latexsym}
\usepackage{amsfonts}
\usepackage{graphicx}
\usepackage{mathrsfs}
\usepackage{epstopdf}
\usepackage{subfigure}

\begin{document}

\title{Nonminimal coupling and inflationary attractors}

\author{Zhu Yi}
\email{yizhu92@hust.edu.cn}
\affiliation{School of Physics, Huazhong University of Science and Technology,
Wuhan, Hubei 430074, China}

\author{Yungui Gong}
\email{yggong@mail.hust.edu.cn}
\affiliation{School of Physics, Huazhong University of Science and Technology,
Wuhan, Hubei 430074, China}


\begin{abstract}
We show explicitly how the T model, E model, and Hilltop inflations are
obtained from the general scalar-tensor theory of gravity with arbitrary conformal factors in the strong coupling limit.
We argue that $\xi$ attractors can give any observables $n_s$ and $r$ by this method.
The existence of attractors imposes a challenge to distinguish different models.
\end{abstract}

\preprint{1608.05922}

\maketitle
\section{Introduction}

Inflation not only solves the flatness, horizon, and monopole problems in the standard big bang cosmology,
but also provides the initial condition for the formation of the large-scale structure \cite{starobinskyfr, guth81, linde83, Albrecht:1982wi}.
The quantum fluctuations of the inflaton lead to metric perturbations and imprint the anisotropy signatures in the
cosmic microwave background radiation.
The Planck measurements of the cosmic microwave background
anisotropies give $n_s=0.968\pm 0.006$ and $r_{0.002}<0.11$ (95\% C.L.) \cite{Adam:2015rua,Ade:2015lrj}.
The central value of $n_s$ suggests the relation $n_s=1-2/N$ with $N=60$, where $N$ is the number of $e$-folds before the end of inflation.
Based on this observation, various parametrizations with $N$ for $n_s$, the inflaton $\phi$, the slow roll parameters $\epsilon$
and $\eta$ were proposed to study the property of inflationary models \cite{Mukhanov:2013tua,Roest:2013fha,Garcia-Bellido:2014gna,Barranco:2014ira,Boubekeur:2014xva,Chiba:2015zpa,Creminelli:2014nqa,Gobbetti:2015cya,Lin:2015fqa}.
The parametrizations of $\epsilon(N)$, $\eta(N)$, $n_s(N)$ or $\phi(N)$ can be compared with the observational data directly and used
to reconstruct the inflationary potentials in the slow roll regime \cite{Lin:2015fqa}.

Many inflationary models have the universal result $n_s=1-2/N$ to the leading order approximation in $1/N$ for large $N$.
In the Einstein frame, the $R^2$ inflation \cite{starobinskyfr} has the effective potential $V(\phi)=V_0[1-\exp(-\sqrt{2/3}\,\phi)]^2$,
so the $R^2$ inflation has the scalar tilt $n_s=1-2/N$ and the tensor to scalar ratio $r=12/N^2$. For the Higgs inflation
with the nonminimal coupling $\xi\phi^2 R$, the same results were obtained in the strong coupling limit $\xi\gg 1$ \cite{Kaiser:1994vs,Bezrukov:2007ep}.
Although the results can be obtained for the coupling constant as small as $\xi> 0.1$ \cite{Kaiser:1994vs}, and the derived $n_s$ and $r$ are
consistent with the Planck observations even when $\xi>0.003$ \cite{Bezrukov:2013fca,Boubekeur:2015xza}, the amplitude of the scalar perturbation
$A_s=2.2\times 10^{-9}$ \cite{Ade:2015lrj} requires that $\xi\approx 49000\sqrt{\lambda}$ \cite{Bezrukov:2007ep}; therefore, the strong coupling
limit $\xi\gg 1$ is needed to satisfy the observational constraints. The universal attractor for $R^2$ inflation and the Higgs inflation was
also derived from more general nonminimal coupling $\xi f(\phi)R$ with the potential $\lambda^2 f^2(\phi)$ for arbitrary function $f(\phi)$
in the strong coupling limit \cite{Kallosh:2013tua}. For the conformal coupling $\xi\phi^2 R$ with $\xi=-1/6$, the kinetic term for the scalar
field $\phi$ in the Einstein frame becomes $\partial\phi^2/(1-\phi^2/6)^2$ and has poles. In terms of the canonical scalar field $\varphi=\sqrt{6}\tanh^{-1}(\phi/\sqrt{6})$,
the potential becomes much flatter, which can be used to discuss non-slow-roll dynamics in the original Jordan frame \cite{Kumar:2015mfa}.
If we take the monomial potential $V(\phi)$, then we get the T model $V(\varphi)=V_0\tanh^{2n}(\varphi/\sqrt{6})$ in the Einstein frame and the universal attractor was also obtained \cite{Kallosh:2013hoa,Kallosh:2013maa}.
The universal attractor was generalized to the $\alpha$ attractors with the same $n_s$ and $r=12\alpha/N^2$ by varying the K\"{a}hler curvature \cite{Kallosh:2013yoa},
in which the kinetic term was generalized to $\partial\phi^2/(1-\phi^2/6\alpha)^2$ and the relation between $\phi$ and the canonical field $\varphi$ becomes
$\phi=\sqrt{6\alpha}\tanh(\varphi/\sqrt{6\alpha})$ \cite{Kallosh:2016gqp}. Inflation due to the existence of the poles in the kinetic term
was discussed in more detail as the generalized pole inflation \cite{Terada:2016nqg}.
Furthermore, these attractors can be described by a general scalar-tensor theory of gravity \cite{Galante:2014ifa}.

In this paper, we show that in the strong coupling limit, any attractor is possible for the nonminimal coupling with an arbitrary coupling function $f(\phi)$.
We also discuss how long the coupling constant $\xi$ takes to reach the attractors and the constraint on the
energy scale of the model from $A_s$. The paper is organized as follows.
In Sec. 2, we discuss the derivation of various attractors. The detailed behavior of the $\xi$ attractors are discussed in Sec. 3. The conclusions are drawn in Sec. 4.

\section{The inflationary attractors}

\subsection{The attractors in scalar-tensor theory}

The action of a general scalar-tensor theory in the Jordan frame is
\begin{equation}
\label{jscten} S=\int d^4x\sqrt{-\tilde{g}}\left[\frac{1}{2}\Omega(\phi)\tilde{R}(\tilde{g})-{1\over
2}\omega(\phi)\tilde{g}^{\mu\nu}
\nabla_{\mu}\phi\nabla_{\nu}\phi-V_J(\phi)\right],
\end{equation}
where $\Omega(\phi)=1+\xi f(\phi)$ with the dimensionless coupling constant $\xi$, $f(\phi)$ is an arbitrary function, and
the scalar field is normalized by the reduced Planck mass $M_{pl}=(8\pi G)^{-1}=1$.
If we take the following conformal transformations:
\begin{gather}
\label{conftransf1}
g_{\mu\nu}=\Omega(\phi){\tilde g}_{\mu\nu},\\
\label{conftransf2}
d\psi^2=\left[\frac{3}{2}\frac{(d\Omega/d\phi)^2}{\Omega^2(\phi)}+\frac{\omega(\phi)}{\Omega(\phi)}\right]d\phi^2,
\end{gather}
then the action (\ref{jscten}) in the Einstein frame becomes
\begin{equation}
\label{escten}
S=\int d^4x\sqrt{-g}\left[\frac{1}{2}R(g)-\frac{1}{2}g^{\mu\nu}\nabla_\mu\psi
\nabla_\nu\psi-U(\psi)\right],
\end{equation}
where $U(\psi)=V_J(\phi)/\Omega^2(\phi)$.

If the conformal factor $\Omega(\phi)$ and the kinetic coupling $\omega(\phi)$ satisfy the condition \cite{Galante:2014ifa}
\begin{equation}\label{coup2}
\omega(\phi)=\frac{1}{4\xi}\frac{(d\Omega(\phi)/d\phi)^2}{\Omega(\phi)},
\end{equation}
then there exists an exact relationship between $\phi$ and $\psi$,
\begin{equation}
\label{psiphirel2}
\psi= \sqrt{\frac{3\alpha}{2}}\ln\Omega(\phi),\quad \Omega(\phi)= e^{\sqrt{2/3\alpha}\,\psi},
\end{equation}
and
\begin{equation}\label{pontrel2}
V_J(\phi)=\Omega^2(\phi)U\left(\sqrt{\frac{3\alpha}{2}}\ln\Omega(\phi)\right),
\end{equation}
where $\alpha=1+(6\xi)^{-1}$. Under the condition \eqref{coup2}, if we take $V_J(\phi)=V_0 [1-\Omega(\phi)]^2$, then we get
\begin{equation}
\label{alphaeq1}
U(\psi)=V_0[1-\exp(-\sqrt{2/3\alpha}\,\psi)]^2,
\end{equation}
and the $\alpha$ attractors \cite{Ferrara:2013rsa,Kallosh:2013yoa},
\begin{equation}
\label{attractor2}
n_s=1-\frac{2}{N},\quad r=\frac{12\alpha}{N^2},
\end{equation}
for small $\alpha$. Note that the above result is independent of the function $\Omega(\phi)$.
Under the condition \eqref{coup2}, if we take
\begin{equation}
\label{tmodel2}
V_J(\phi)=V_0\Omega^2\left[1-\Omega^{-1}(\phi)\right]^{2n},
\end{equation}
then we get the general E-model potential in the Einstein frame,
\begin{equation}
\label{tmodel1}
U(\psi)=V_0\left[1-\exp\left(-\sqrt{\frac{2}{3\alpha}}\,\psi\right)\right]^{2n}.
\end{equation}
The $\alpha$ attractors \eqref{attractor2} are also obtained to the leading order of $1/N$ for small $\alpha$ \cite{Carrasco:2015rva}.
Under the condition \eqref{coup2}, if we take
\begin{equation}
\label{tmodel2}
V_J(\phi)=V_0\Omega^2\left(\frac{1-\Omega(\phi)}{1+\Omega(\phi)}\right)^{2n},
\end{equation}
then we get the general T-model potential in the Einstein frame,
\begin{equation}
\label{tmodel1}
U(\psi)=V_0\tanh^{2n}\frac{\psi}{\sqrt{6\alpha}}.
\end{equation}
The $\alpha$ attractors \eqref{attractor2} are obtained by the leading order of $1/N$ for small $\alpha$ \cite{Kallosh:2013yoa}.
Because when $\alpha\ll 1$, the potentials of the T model and E model  have the same asymptotic behavior
\begin{equation}
\label{univvphi}
U(\psi)=V_0\left[1-c \exp\left({-\sqrt{\frac{2}{3\alpha}}\,\psi}\right)\right],
\end{equation}
so both T model and E model have the same $\alpha$ attractors \cite{Kallosh:2013hoa}, and the $\alpha$ attractors can be
achieved from several broad classes
of models by using the conformal transformation \eqref{psiphirel2}.

To the leading order of slow-roll approximation, using the relationship \cite{Lin:2015fqa}
\begin{equation}
\label{pontreeq1}
r \approx 16\epsilon\approx 8 \frac{d\ln V}{dN},
\end{equation}
the reconstructed potential with the $\alpha$ attractors is
\begin{equation}
\label{recponteq2}
U(\psi)=V_0\exp\left[{-\frac{3\alpha}{2}e^{-\sqrt{\frac{2}{3\alpha}}(\psi-\psi_0)}}\right].
\end{equation}
When $\alpha\ll 1$, the above potential reduces to the potential \eqref{univvphi}.

If $\Omega(\phi)=1$, then $d\psi^2=\omega(\phi)d\phi^2$. By choosing the function $\omega(\phi)$, we may obtain the
$\alpha$ attractors
and the Hilltop inflation.
For example, the T model can be obtained by choosing \cite{Galante:2014ifa}
\begin{equation}
\label{poletmodeq1}
\omega(\phi)=\frac{1}{(1-\phi^2/6\alpha)^2},
\end{equation}
and $V_J(\phi)=V_0\phi^{2n}$.
The E model is obtained by choosing $\omega(\phi)=3\alpha/(2\phi^2)$ and $V_J(\phi)=V_0(1-c\phi)^{2n}$.
If $\omega(\phi)$ has the pole of order $p\neq 2$ \cite{Galante:2014ifa},
\begin{equation}\label{polehilltopeq1}
\omega(\phi)=\frac{a_p}{\phi^p},
\end{equation}
then we can obtain the Hilltop potential
\begin{equation}\label{hilltopeq2}
U(\psi)=V_0\left[1-\left(\frac{\psi}{\mu}\right)^{n}\right],
\end{equation}
with $p=2-2/n<2$ by choosing $V_J(\phi)=V_0(1-c\phi)$.

If the contribution from $\omega(\phi)$ is negligible,  i.e., if the conformal factor satisfies the condition
\begin{equation}
\label{strongcoup1}
\Omega(\phi)\ll \frac{3(d\Omega(\phi)/d\phi)^2}{2\omega(\phi)},
\end{equation}
then the relationship between $\psi$ and $\phi$ in Eq. \eqref{psiphirel2} holds approximately, and $\alpha=1$. Therefore,
\begin{equation}
\label{psiphirel1}
\psi\approx \sqrt{\frac{3}{2}}\ln\Omega(\phi),\quad \Omega(\phi)\approx e^{\sqrt{2/3}\,\psi},
\end{equation}
and
\begin{equation}\label{pontrel1}
V_J(\phi)\approx \Omega^2(\phi)U\left(\sqrt{\frac{3}{2}}\ln\Omega(\phi)\right).
\end{equation}
For the scalar-tensor theory of gravity with $\Omega(\phi)=1+\xi f(\phi)$, $\omega(\phi)=1$ and $V_J(\phi)=V_0[1-\Omega(\phi)]^2=\xi^2 V_0 f^2(\phi)$ \cite{Kallosh:2013tua},
under the strong coupling limit $\xi\gg 1$, we get the potential \eqref{alphaeq1} and the universal attractor \eqref{attractor2} with $\alpha=1$ independent of the choices of the
arbitrary function $f(\phi)$. The Higgs inflation with the nonminimal coupling $\xi\phi^2 R$ is the special case with $f(\phi)=\phi^2$.

\subsection{The attractors in nonlinear f(R) gravity}

For the nonlinear $f(R)$ gravity, the action is
\begin{equation}\label{frgravity}
S=\int d^4x \frac{1}{2}\sqrt{-\tilde{g}}f(\tilde{R})=\frac{1}{2}\int d^4x \sqrt{-\tilde{g}}[f'(\phi)\tilde{R}-\phi f'(\phi)+f(\phi)],
\end{equation}
so it can be thought of as a scalar-tensor theory of gravity with the conformal factor $\Omega(\phi)=f'(\phi)$
and the scalar potential $V_J(\phi)=[\phi f'(\phi)-f(\phi)]/2$, where $f'(\phi)=df(\phi)/d\phi$.
After the conformal transformation,
\begin{equation}
\label{frconfeq2}
g_{\mu\nu}=f'(\phi)\tilde{g}_{\mu\nu},\quad f'(\phi)=c_0\exp\left(\sqrt{\frac23}\psi(\phi)\right),
\end{equation}
the potential in Einstein frame becomes
\begin{equation}\label{efrpont1}
U(\psi)=\frac{1}{2c_0^2}\exp\left(-2\sqrt{\frac{2}{3}}\,\psi\right)\left(\phi(\psi) f'[\phi(\psi)]-f[\phi(\psi)]\right),
\end{equation}
and the relationship between $\phi$ and $\psi$ is
\begin{equation}
\label{phipsirel1}
  \phi(\psi)=\sqrt{6}c_0\partial_\psi \left[U(\psi)\exp\left(2\sqrt{\frac23}\psi\right)\right] \exp\left(-\sqrt{\frac23}\psi\right),
\end{equation}
where $c_0$ is an arbitrary constant.
For a given potential $U(\psi)$,  we can derive the relationship between $\phi$ and $\psi$ from Eq. \eqref{phipsirel1}
and then find out the corresponding function $f(R)$ from Eq. \eqref{frconfeq2}.
For example, if we choose the potential
\begin{equation}
\label{r2pot12}
  U(\psi)=V_0\left[1-c \exp\left(-\sqrt{\frac23}\psi\right)\right]^2,
\end{equation}
then we get the relationship between $\phi$ and $\psi$ from Eq. \eqref{phipsirel1} as
\begin{equation}
\label{phipsirel2}
  \phi=4c_0V_0\left[\exp\left(\sqrt{\frac23}\psi\right)-c\right],
\end{equation}
Substituting Eq. \eqref{phipsirel2} into Eq. \eqref{frconfeq2} and choosing $c_0=1/c$, we get
\begin{equation}
  f'(\phi)=\frac{\phi}{4V_0}+1.
\end{equation}
Therefore, the Starobinsky model $f(R)=R+R^2/(8V_0)$ has the effective potential \eqref{r2pot12} with the universal attractors $n_s=1-2/N$ and $r=12/N^2$.

For the potential \eqref{univvphi}, we get
\begin{equation}
  \phi=4V_0c_0\exp\left(\sqrt{\frac23}\psi\right)\left[1-c\left(1-\sqrt{\frac{1}{4\alpha}}\right)\exp\left(-\sqrt{\frac{2}{3\alpha}}\psi\right)\right].
\end{equation}
when $\alpha\ll1$, we have
\begin{gather}
  \ln\left(\frac{\phi}{4V_0c_0}\right)\approx\sqrt{\frac23}\psi-c\left(1-\sqrt{\frac{1}{4\alpha}}\right)\exp\left(-\sqrt{\frac{2}{3\alpha}}\psi\right),\\
  \label{psiphirelalpha}
  \psi\approx\sqrt{\frac32}\ln\left(\frac{\phi}{4V_0c_0}\right)+\sqrt{\frac32}c\left(1-\sqrt{\frac{1}{4\alpha}}\right)\left(\frac{\phi}{4V_0c_0}\right)^{-1/\sqrt{\alpha}}.
\end{gather}
Combining Eqs. \eqref{frconfeq2} and \eqref{psiphirelalpha} and choosing $c_0=(8V_0/c)^{\sqrt{\alpha}}/(4V_0)$, we get
\begin{equation}
\label{fphide}
\begin{split}
f'(\phi)&=\frac{\phi}{4V_0}\exp\left[4V_0\left(2-\sqrt{\frac{1}{\alpha}}\right)\phi^{-1/\sqrt{\alpha}}\right]\\
&\approx \frac{\phi}{4V_0}+\left(2-\sqrt{\frac{1}{\alpha}}\right)\phi^{1-1/\sqrt{\alpha}}.
\end{split}
\end{equation}
Solving Eq. \eqref{fphide} and neglecting the integration constant which corresponds to the cosmological constant,
we obtain the $f(R)$ model with the $\alpha$ attractors \eqref{attractor2},
\begin{equation}
\label{frgamma}
\begin{split}
  f(R)&=-\frac{\sqrt{\alpha}}{4V_0}\left[4V_0\left(\frac{1}{\sqrt{\alpha}}-2\right)\right]^{2\sqrt{\alpha}}\gamma
  \left[-2\sqrt{\alpha},4V_0\left(\frac{1}{\sqrt{\alpha}}-2\right)R^{-\frac{1}{\sqrt{\alpha}}}\right]\\
  &\approx R^{2-\frac{1}{\sqrt{\alpha}}}+\frac{R^2}{8V_0},
\end{split}
\end{equation}
where the Gamma function $\gamma(a,x)=\int_0^x e^{-t}t^{a-1}dt$.
Interestingly, although the derivation of the $f(R)$ model \eqref{frgamma} is based on the condition $\alpha\ll 1$,
the result extends to recover the Starobinsky model when $\alpha=1$.

From the above discussions, we see that the potential \eqref{alphaeq1} and the $\alpha$ attractors \eqref{attractor2}
can be obtained from either $f(R)$ gravity or the scalar-tensor theory of gravity.
We will show that we can obtain whatever attractors we want by using the above transformations.

\section{The General $\xi$ attractors}

As discussed in the previous section, under the limit \eqref{strongcoup1},
we can get any attractor behavior associated with the arbitrary potential $U(\psi)$ if we assume $V_J(\phi)$ takes the form defined in Eq. \eqref{pontrel1}.
To get the E-model attractor with the potential \cite{Kallosh:2013maa,Carrasco:2015rva},
\begin{equation}\label{emodeleq1}
U(\psi)=V_0\left[1-\exp\left(-\sqrt{\frac{2}{3\alpha}}\,\psi\right)\right]^{2n},
\end{equation}
we take
\begin{equation}
\label{emodeleq2}
V_J(\phi)=V_0\Omega^2(\phi)\left[1-\Omega^{-1/\sqrt{\alpha}}(\phi)\right]^{2n}.
\end{equation}
 The spectral index $n_s$ and the tensor to scalar ratio $r$ are
\begin{gather}
\label{emodnseq1}
n_s=1+\frac{8 n }{3 \alpha \left[ g(N,n,\alpha)+1\right]}-\frac{8  n (n+1)}{3\alpha  \left[ g(N,n,\alpha)+1\right]^2},\\
\label{emodreq1}
r=\frac{64  n^2}{3 \alpha \left[g(N,n,\alpha)+1\right]^2},
\end{gather}
where
\begin{equation}
  \label{emodg1}
g(N,n,\alpha)=W_{-1}\left[-\left(\frac{2n}{\sqrt{3\alpha }}+1\right) \exp \left(\frac{-4nN}{3\alpha}-\frac{2n}{\sqrt{3\alpha}}-1\right)\right],
\end{equation}
for $n>1$ and $n/{3(2n-1)}<\alpha<{4n^2}/{3(n-1)^2}$, or $1/3<n<1$ and $\alpha>{4n^2}/{3(3n-1)^2}$;
\begin{equation}
\label{emodg2}
   g(N,n,\alpha)=W_{-1}\left[-\left(\frac{2u}{3 \alpha }-\frac{2 n}{3 \alpha }+1\right)
   \exp\left({-1-\frac{2 u+2 n (2N-1)}{3 \alpha }}\right)\right],
\end{equation}
for $n>1$ and $\alpha>{4n^2}/{3(n-1)^2}$;
\begin{equation}
\label{emodg3}
 g(N,n,\alpha)=W_{-1}\left[-\left(\frac{2 n}{3 \alpha }+\frac{2 v}{3 \alpha }+1\right)
 \exp\left({-1-\frac{2v+2n (2N+1)}{3 \alpha }}\right)\right],
\end{equation}
for other cases; $u=\sqrt{6 \alpha  n^2+n^2-3 \alpha  n}$, $v=\sqrt{n (3 \alpha -6 \alpha  n+n)}\,$ and $W_{-1}$ is the lower branch of the Lambert $W$ function.
When $\alpha\ll 1$, Eq. \eqref{emodg3} can be approximated as
\begin{equation}
g(N,n,\alpha)=-\frac{4 n (N+1)}{3 \alpha }-\ln (N+1),
\end{equation}
so $n_s$ and $r$ become \cite{Carrasco:2015rva}
\begin{gather}
 \label{emodnseq4}
  n_s=1-\frac{2}{1+N-3\alpha \ln(1+N)/(4n)}\approx 1-\frac{2}{N},\\
   \label{emodreq4}
  r=\frac{12\alpha}{(1+N)^2+3\alpha(1+N)\ln(1+N)/(2n)}\approx \frac{12\alpha}{N^2}.
\end{gather}
The deviation from the above attractor for $\alpha\sim 1$ is bigger if $n\sim 1$ is smaller.

\begin{figure}[htbp]
\centering
\includegraphics[width=0.6\textwidth]{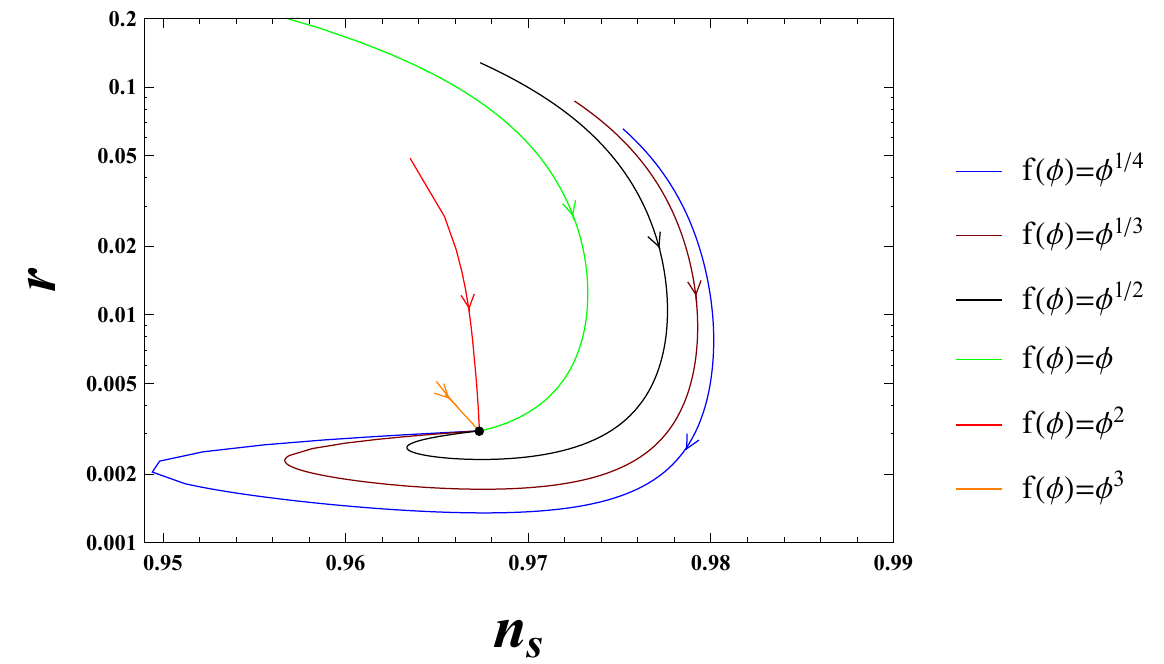}
\caption{The numerical results of $n_s$ and $r$ for the scalar-tensor theory with the potential \eqref{emodeleq2}. We take
$\omega(\phi)=1$, $n=2$, $\alpha=1$, the power-law functions $f(\phi)=\phi^k$ with
$k=1/4$, 1/3, 1/2, 1, 2, and 3, and $N=60$. The coupling constant $\xi$ increases along the direction of the arrow in the plot.
The E-model attractors \eqref{emodnseq1} and  \eqref{emodreq1} are reached if $\xi\gtrsim 100$.}
\label{fig:Ensr}
\end{figure}

To show that the attractors \eqref{emodnseq1} and \eqref{emodreq1} can be reached for an arbitrary conformal factor $\Omega(\phi)=1+\xi f(\phi)$,
as an example, we take $\omega(\phi)=1$, $n=2$, $\alpha=1$, and the power-law functions $f(\phi)=\phi^k$ with
$k=1/4$, 1/3, 1/2, 1, 2, and 3. We vary the coupling constant $\xi$ and choose $N=60$ to calculate $n_s$ and $r$ for
the models with the potential \eqref{emodeleq2}, the results are shown in Fig. \ref{fig:Ensr}.
From Fig. \ref{fig:Ensr}, we see that the E-model attractor is reached when $\xi\gtrsim 100$.
From Eqs. \eqref{emodnseq1},  \eqref{emodreq1},  and  \eqref{emodg1}, we get the attractors
$n_s=0.9673$ and $r=0.0031$ for $N=60$. The amplitude of the scalar perturbation
 \begin{equation}
 \label{emodaseq}
   As\simeq\frac{3 \left\{W_{-1}\left[-\left(1+\frac{4}{\sqrt{3}}\right) e^{-\frac{4}{3} \left(2 N+\sqrt{3}\right)-1}\right]+1\right\}^6}{16 W_{-1}\left[-\left(1+\frac{4}{\sqrt{3}}\right) e^{-\frac{4}{3}  \left(2 N+\sqrt{3}\right)-1}\right]^4}V_0 =5058.38 V_0 .
 \end{equation}
Applying the observational result $A_s=2.2\times 10^{-9}$ \cite{Ade:2015lrj}, we get the energy scale $V_0$ for the E model
$V_0=4.35 \times 10^{-13}$.

To get the T-model attractor with the potential \cite{Kallosh:2013hoa,Kallosh:2013maa},
\begin{equation}\label{tmodeleq1}
U(\psi)=V_0\tanh^{2n}\left(\frac{\psi}{\sqrt{6\alpha}}\right),
\end{equation}
we take
\begin{equation}\label{tmodeleq2}
V_J=V_0\Omega^2(\phi)\left[\frac{1-\Omega^{2/\sqrt{\alpha}}(\phi)}{1+\Omega^{2/\sqrt{\alpha}}(\phi)}\right]^{2n}.
\end{equation}
The spectral index $n_s$ and the tensor to scalar ratio $r$ are \cite{Kallosh:2013yoa}
\begin{gather}
 \label{tmodnseq1}
n_s=1-\frac2N+\frac{2 N\sqrt{12  n^2/\alpha+9}-6 n (N-1)}{N \left[2N \sqrt{12 n^2/\alpha+9}+n \left(4 N^2/\alpha+3\right)\right]},\\
\label{tmodreq1}
r=\frac{48 n}{2 N \sqrt{12 n^2/\alpha+9}+n \left(4  N^2/\alpha+3\right)},
\end{gather}
for $n>1$ and $(4n^2-2n\sqrt{4n^2-1})/3<\alpha<{4n^2}/{3(n^2-1)}$, $1/\sqrt{3}<n<1$ and $\alpha>(4n^2-2n\sqrt{4n^2-1})/3$ or  $1/3<n<1/\sqrt{3}$ and $\alpha>4n^2/3(9 n^2-1)$;
  \begin{gather}
   \label{tmodnseq2}
  n_s=1-\frac2N+\frac{8 n \left[(N-1) \sqrt{9 \alpha ^2+24 \alpha  n^2+4 n^2}+6 \alpha  n-(3 \alpha +2) n N+2 n\right]}{N \left[\left(\sqrt{9 \alpha ^2+4 (6 \alpha +1) n^2}+4 n N-2 n\right)^2-9 \alpha ^2\right]},\\
  \label{tmodreq2}
  r=\frac{192 \alpha  n^2}{\left(\sqrt{9 \alpha ^2+4 (6 \alpha +1) n^2}-2 n+4 n N\right)^2-9 \alpha ^2},
\end{gather}
for $n>1$ and $\alpha>{4n^2}/{3(n^2-1)}$;
\begin{gather}
\label{tmodnseq3}
n_s=1-\frac2N+\frac{8 n \left[(N+1) \sqrt{9 \alpha ^2+(4-24 \alpha ) n^2}-6n \alpha -(3 \alpha -2)n N+2n\right]}{N \left[\left(\sqrt{9 \alpha ^2+(4-24 \alpha ) n^2}+4 n N+2n\right)^2-9 \alpha ^2\right]},\\
\label{tmodreq3}
r=\frac{192 \alpha  n^2}{\left[\sqrt{9 \alpha ^2+(4-24 \alpha ) n^2}+n (4 N+2)\right]^2-9 \alpha ^2},
\end{gather}
for other cases.
If  $\alpha \ll 1$, Eqs. \eqref{tmodnseq3} and \eqref{tmodreq3}  become $n_s=1-2/N+{2}/{(N^2+N)}$ and $r={12\alpha}/{ (1+N)^2}$.

\begin{figure}[htbp]
\centering
\includegraphics[width=0.6\textwidth]{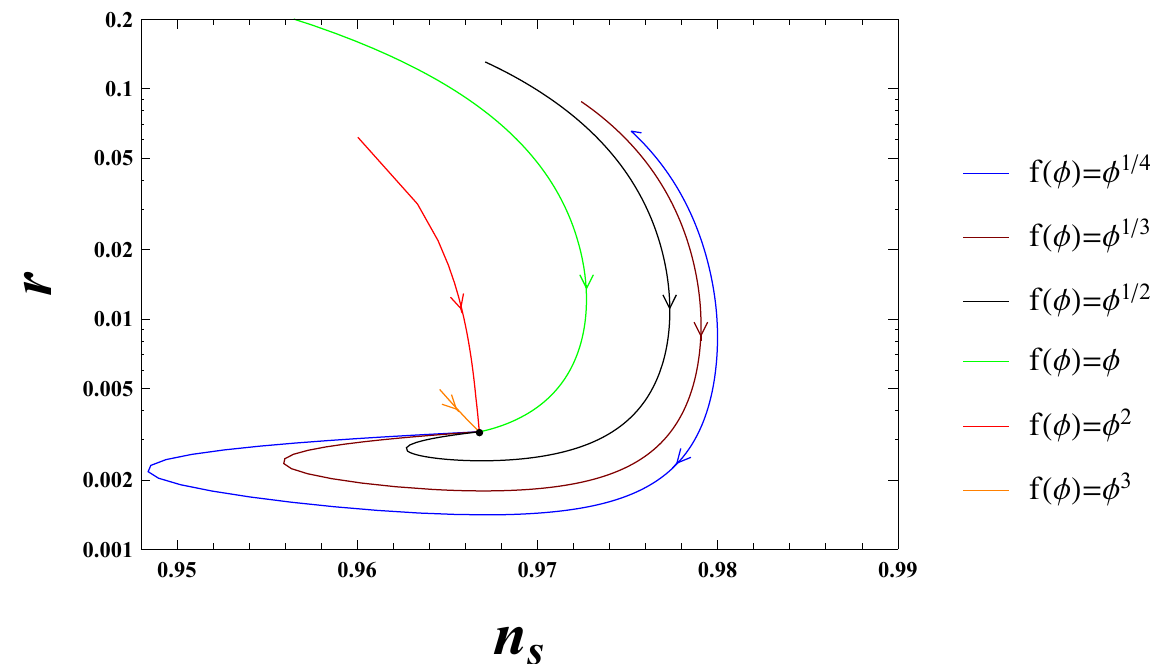}
\caption{The numerical results of $n_s$ and $r$ for the scalar-tensor theory with the potential \eqref{tmodeleq2}. We take
$\omega(\phi)=1$, $n=2$, $\alpha=1$, the power-law functions $f(\phi)=\phi^k$ with
$k=1/4$, 1/3, 1/2, 1, 2, and 3, and $N=60$. The coupling constant $\xi$ increases along the direction of the arrow in the plot.
The T-model attractors \eqref{tmodnseq1} and  \eqref{tmodreq1} are reached if $\xi\gtrsim 100$.}
\label{fig:Tnsr}
\end{figure}

To calculate $n_s$ and $r$ explicitly for the potential \eqref{tmodeleq2},
we take $\omega(\phi)=1$, $n=2$, $\alpha=1$, and the power-law functions $f(\phi)=\phi^k$ with
$k=1/4$, 1/3, 1/2, 1, 2, and 3. We choose $N=60$ and vary the coupling constant $\xi$;
the results are shown in Fig. \ref{fig:Tnsr}.
From Eqs. \eqref{tmodnseq1} and \eqref{tmodreq1}, we get the attractors $n_s=0.9668$ and $r=0.0032$.
From Fig. \ref{fig:Tnsr}, we find that the T-model attractor is reached when $\xi\gtrsim 100$.
Using the observational constraint on $A_s$ \cite{Ade:2015lrj}, we get the energy scale of the potential  $V_0=4.55 \times10^{-13}$.

To get the hilltop attractor with the potential,
\begin{equation}
\label{hilltopeq1}
U(\psi)=V_0\left[1-\left(\frac{\psi}{\mu}\right)^n\right],
\end{equation}
we take
\begin{equation}
\label{hilltopeq2}
V_J(\phi)=V_0\Omega^2(\phi)\left[1-\left(\frac{\sqrt{3/2}\,\ln\Omega(\phi)}{\mu}\right)^n\right].
\end{equation}
Here we consider the cases of $n>2$ and $\mu<1$ only. The spectral index $n_s$  and the tensor to scalar ratio $r$ are
\begin{gather}
\label{hilltopnseq}
n_s=1-\frac{ 2(n-1)}{(n-2)N+n-1}-\frac{n \left[(5 n-2) p(N,n,\mu)^{\frac{n}{n-2}}-2 (n-1)\right]}{\mu ^2 p(N,n,\mu) \left[-2 p(N,n,\mu)^{\frac{n}{n-2}}+p(N,n,\mu)^{\frac{2 n}{n-2}}+1\right]},\\
\label{hilltopreq}
r=\frac{8 n^2}{\mu ^2 \left[p(N,n,\mu)^{\frac{n-1}{n-2}}-p(N,n,\mu)^{-\frac{1}{n-2}}\right]^2},
\end{gather}
where
\begin{equation}
  p(N,n,\mu)=\frac{n [(n-2)N+n-1]}{\mu ^2}.
\end{equation}
In the large $N$ limit, Eqs. \eqref{hilltopnseq} and \eqref{hilltopreq}  become \cite{Boubekeur:2005zm}
\begin{equation}
  n_s\approx1-\frac{ 2(n-1)}{(n-2)N}-\frac{(5n-2)n}{\mu^2}\left[\frac{\mu^2}{n(n-2)N}\right]^{(2n-2)/(n-2)},
\end{equation}
\begin{equation}
  r\approx\frac{8 n^2}{\mu^2}\left[\frac{\mu^2}{n(n-2)N}\right]^{(2n-2)/(n-2)}.
\end{equation}

\begin{figure}[htbp]
\centering
\subfigure[The dependence of $n_s$ on $\xi$]{
\includegraphics[width=0.36\columnwidth]{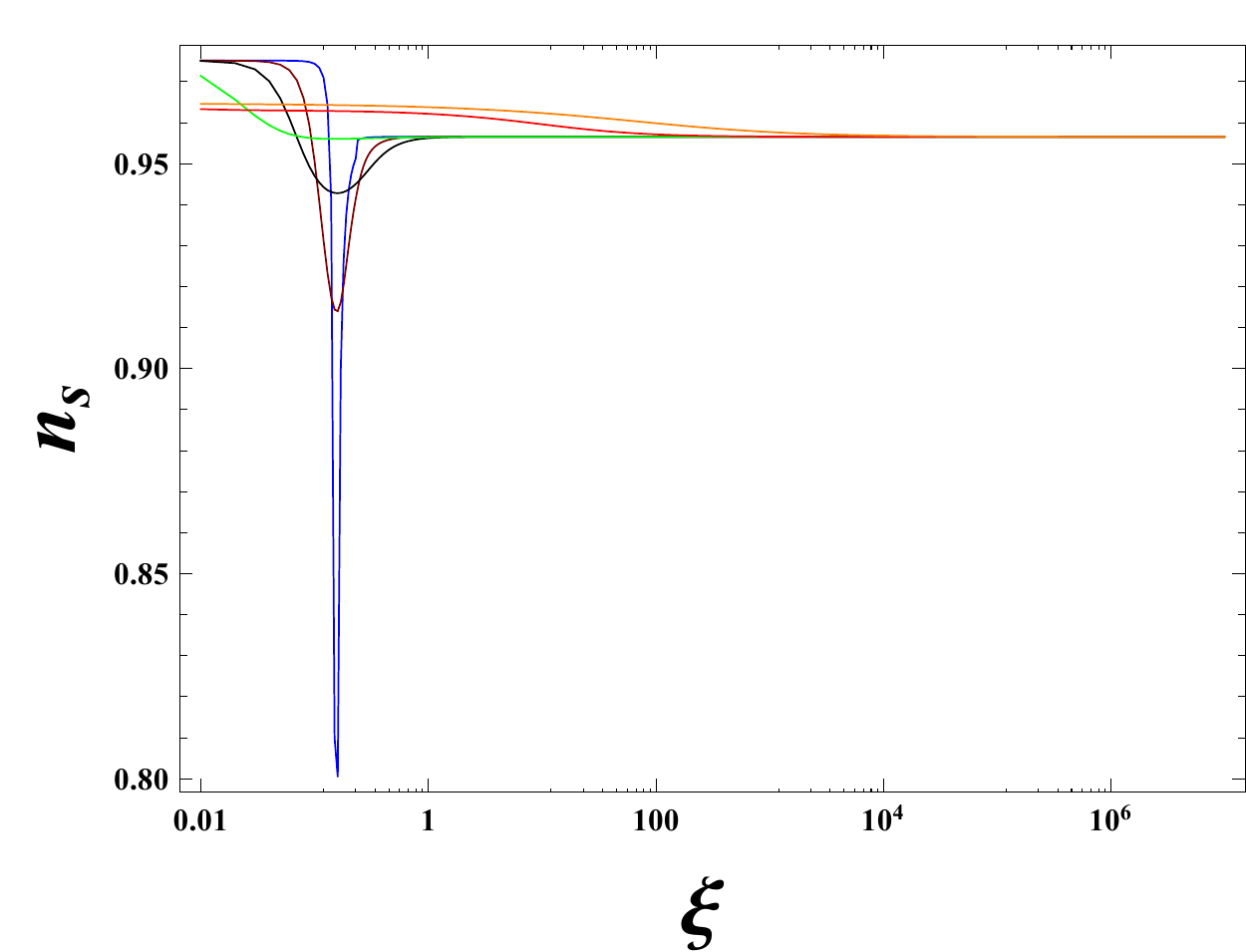}
\label{fig:hns}
}
\subfigure[The dependence of $r$ on $\xi$]{\label{fig:hr}
 \includegraphics[width=0.45\columnwidth]{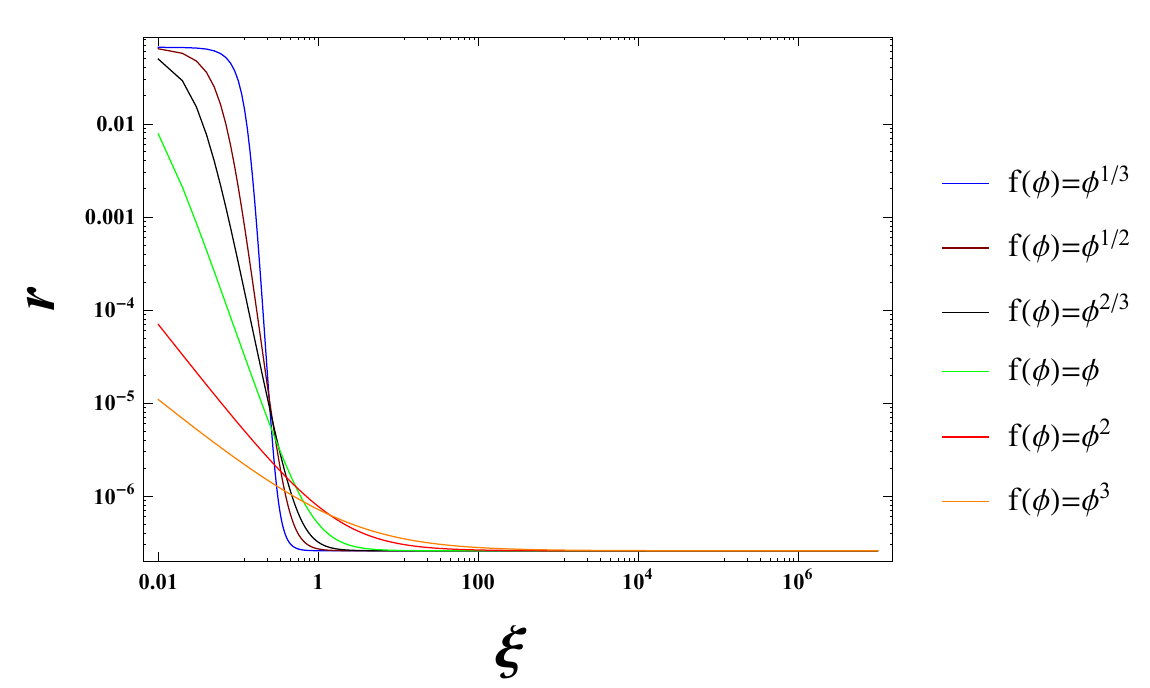}
}
\caption{The dependence of $n_s$ and $r$ on $\xi$ for the scalar-tensor theory with the potential \eqref{hilltopeq2}. We take
$\omega(\phi)=1$, $n=12$, $\mu=1/3$, the power-law functions $f(\phi)=\phi^k$ with
$k=1/3$, 1/2, 2/3, 1, 2 and 3, and $N=60$.}
\label{fig:h:nsr}
\end{figure}

\begin{figure}[htbp]
\centering
\includegraphics[width=0.6\textwidth]{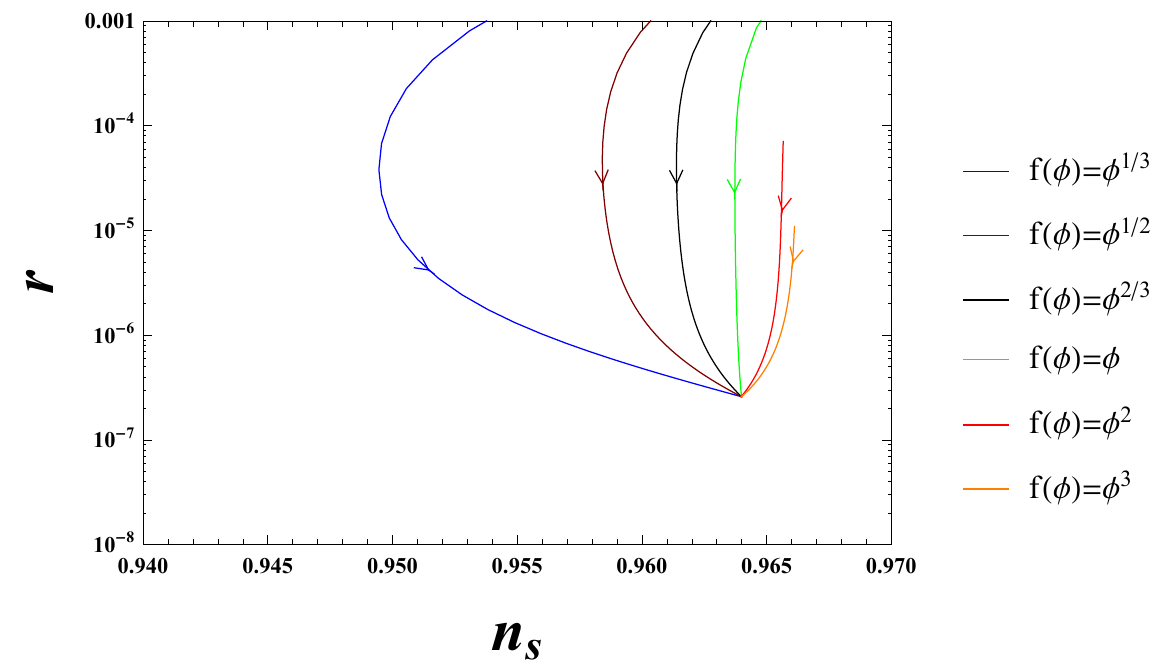}
\caption{ The numerical results of $(n_s,\ r)$ for the scalar-tensor theory with the potential \eqref{hilltopeq2}.
The coupling constant $\xi$ increases along the direction of the arrow in the plot.}
\label{fig:hnsr}
\end{figure}

For the potential \eqref{hilltopeq2}, we choose $\omega(\phi)=1$ and the power-law functions $f(\phi)=\phi^{k}$ with $k=1/3$, 1/2, 2/3, 1, 2 and 3.
We calculate $n_s$ and $r$ for the cases $n=12$ and $\mu=1/3$, and $N$ is taken to be $N=60$. We show the dependence of $n_s$ and $r$ on the coupling
constant $\xi$ in Fig. \ref{fig:h:nsr} and the results for $n_s$ and $r$ in Fig. \ref{fig:hnsr}.
From Eqs. \eqref{hilltopnseq} and \eqref{hilltopreq}, we get the Hilltop attractor $n_s=0.9640$ and $r=2.6\times10^{-7}$. The observational constraint
on $A_s$ gives the energy scale of the potential $ V_0=3.56\times 10^{-17}$. Figures \ref{fig:h:nsr} and \ref{fig:hnsr} tell us that
the attractor is reached if $\xi\gtrsim 1$ for $k\le 1$. For the model with $f(\phi)=\phi^3$, we obtain the attractor if $\xi\gtrsim 10^3$.

\begin{figure}[htbp]
\centering
\includegraphics[width=0.6\textwidth]{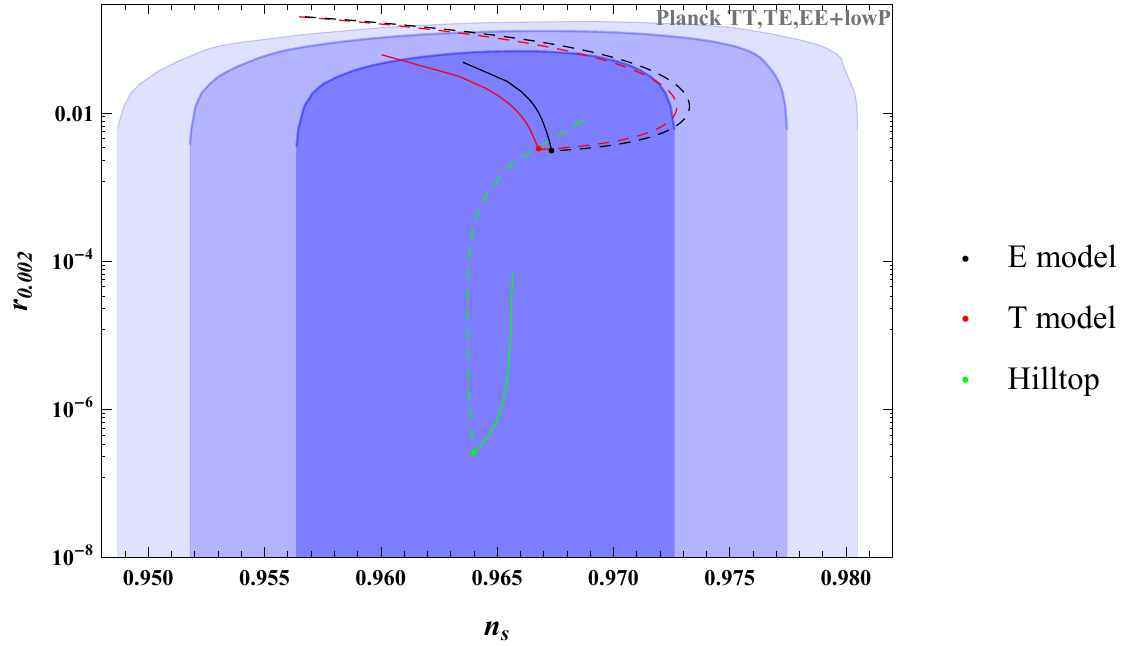}
\caption{The marginalized $68\%$, $95\%$, and $99.8\%$ C.L. contours for $n_s$ and $r_{0.002}$ from Planck 2015 data
and the attractors for T, E, and Hilltop models. The dashed lines are for the coupling function $f(\phi)=\phi$, and the solid lines  represent $f(\phi)=\phi^2$.}
\label{ETH}
\end{figure}

To compare the results with the observations, we replot the $n_s$-$r$ graph
for the three models along with the observational constraints from Planck 2015 data \cite{Ade:2015lrj} in Fig. \ref{ETH},
we show the attractors for the coupling functions $f(\phi)=\phi$ and $f(\phi)=\phi^2$ only for simplicity.
We see that all the three models are consistent with the observational results.

\section{Conclusions}

The T model, E model, and the general potential \eqref{recponteq2} all give
the $\alpha$ attractors, and these models can be constructed in supergravity. The universal
behaviors are due to the hyperbolic geometry of the moduli space and the flatness of the K\"{a}hler potential in the inflaton direction \cite{Carrasco:2015rva}.
Under the conformal transformation, the potentials in the Jordan and Einstein frames have the relationship $U(\psi)=V_J(\phi)/\Omega^2(\phi)$.
For any conformal factor $\Omega(\phi)$, we can always choose the corresponding potential $V_J(\phi)$ in the Jordan frame
so that we have the same potential $U(\psi)$ in the Einstein frame. Therefore, those $\Omega(\phi)$ and $V_J(\phi)$ give the same $\xi$ attractor.
Furthermore, the attractor can be obtained from $f(R)$ theory.
By this method, we may get any attractor we want. In particular, we show explicitly how the T-model, E-model and Hilltop inflations are
obtained for an arbitrary conformal factor $\Omega(\phi)$ in the strong coupling limit. These results further support that
different models give the same observables $n_s$ and $r$, so the existence of $\xi$ attractors imposes a challenge to distinguish different models.

\begin{acknowledgments}

This research was supported in part by the Natural Science
Foundation of China under Grant No. 11475065, and
the Program for New Century Excellent Talents in University under Grant No. NCET-12-0205.

\end{acknowledgments}


\end{document}